\documentclass[]{proarticle}
\usepackage{multicol}
\usepackage{graphicx}
\usepackage{amssymb}
\usepackage{epsfig}

\usepackage{bm}
\usepackage{amsmath}
\usepackage{amsfonts}
\usepackage{slashbox}
\usepackage{comment}

\newcommand{\be}{\begin{equation}}
\newcommand{\ee}{\end{equation}}
\newcommand{\bea}{\begin{eqnarray}}
\newcommand{\eea}{\end{eqnarray}}
\newcommand{\beaa}{\begin{eqnarray*}}
\newcommand{\eeaa}{\end{eqnarray*}}

\newcommand{\nn}{\nonumber \\}
\newcommand{\e}{\mathrm{e}}

\begin{document}
\title{New model of massive spin two field}
\author{Shin'ichi Nojiri}
\maketitle
\address{
Department of Physics, Nagoya University, Nagoya 464-8602, Japan \\
$\&$ Kobayashi-Maskawa Institute for the Origin of Particles and
the Universe, Nagoya University, Nagoya 464-8602, Japan .}
\eads{nojiri@gravity.phys.nagoya-u.ac.jp} 
\begin{abstract}

We propose a new ghost-free model describing massive spin two field.\footnote{
This report is based on the collaborations with Y. Ohara and S. Akagi 
\cite{Ohara:2014vua,Ohara:2014yta,Akagi:2014iea}.
} 
This model consists of a kinetic term and interaction terms without derivative. 
We report on the properties of this model, especially we consider what could happen 
when this model couples with gravity. 
Although the model does not generate any ghost on the Minkowski space-time, it is not so clear 
whether or not this property is preserved even on curved space-time. 
In fact, Buchbinder et al. have found that the ghost appears even in the Fierz-Pauli theory 
on curved space-time if we do not include non-minimal coupling terms. 
We report on the model with interactions on curved space-time and show that we can construct a model 
without ghost by including non-minimal coupling terms.
\end{abstract}


\begin{multicols}{2}

\section{Introduction}

Recently there have been much progress in the study of massive gravity 
\cite{deRham:2010ik,deRham:2010kj} and bigravity 
\cite{Hassan:2011hr,Hassan:2011vm,Hassan:2011zd}, with which motivated, 
we propose a new kind of model for massive spin-two field and investigate the properties.
There are several motivations why we study this model: 1. As we discuss, the condensation of the 
symmetric tensor may break the supersymmetry. 2. The condensation may also explain the 
accelerating expansion of the present universe. 3. The particles corresponding to the massive spin two 
field may be a candidate of the dark matter. 

The free theory of the massive spin two field, which is called massive gravity, was proposed three the fourth 
century ago, by Fierz and Pauli \cite{Fierz:1939ix}. The Lagrangian density for the massless spin-two field 
(graviton) $h_{\mu\nu}$ is given by linearizing the Einstein-Hilbert Lagrangian,  
\begin{align}
\label{T1}
\mathcal{L}_0 = &- \frac{1}{2}\partial_\lambda h_{\mu\nu}\partial^\lambda h^{\mu\nu}
+ \partial_\lambda h^\lambda_{\ \mu}\partial_\nu h^{\mu\nu} 
 - \partial^\mu h_{\mu\nu}\partial^\nu h \nn
& + \frac{1}{2}\partial_\lambda h\partial^\lambda h\, .
\end{align}
Here $h\equiv h^\mu_{\ \mu}$. 
On the other hand, the Lagrangian density of the massive graviton with mass $m$ has the following form:
\be
\label{T2}
\mathcal{L}_m = \mathcal{L}_0 - \frac{m^2}{2}\left(h_{\mu\nu}h^{\mu\nu} - h^2\right)\, ,
\ee
which is called the Fierz-Pauli Lagrangian density. 

We should note that the massless graviton has 2 degrees of freedom corresponding to the helicity but 
the massive graviton has five degrees of freedom because the representation of the spin two states have five 
degrees of freedom (as known in quantum mechanics, the representation of the general spin $s$ states is 
$2s+1$ dimensional). 
The Hamiltonian analysis surely tells that the massive gravity has five propagating degrees of freedom 
in four dimensions. 
In $m=0$ case, $h_{0i}$ and $h_{00}$ play the role of the Lagrange multipliers and give 4 constraints, 
which are the first class constraints associated with the gauge symmetry corresponding to the general 
covariance. 
In four dimensions, $h_{ij}$ and the conjugate momenta $\pi_{ij}$ have 6 components, respectively, which give 12 
dimensional phase space. 
Because of the 4 constraints and 4 gauge invariances, the phase space reduces to 4 dimensional one, 
which corresponds to the two polarizations, that is, helicities of massless graviton. 
On the other hand, in case $m\neq 0$, $h_{0i}$'s are no longer the Lagrange multipliers but the equation given 
by the variation of $h_{0i}$ can be solved with respect to $h_{0i}$. 
Even in case $m\neq 0$, $h_{00}$ plays the role of the Lagrange multiplier and gives a single constraint. 
Because we also obtain a secondary constraint, there are two second class constraints and in four dimensions, 
we obtain 10 degrees of freedom because $12$ dimensional phase space minus 2 constraints are equal to the 
10 physical degrees of freedom, which corresponds to the $5$ polarizations of the massive graviton and their 
conjugate momenta. 

Although we obtain a consistent free field theory of the massive graviton, if we include the interaction, 
there seems to appear always a ghost scalar field called the Boulware-Deser ghost 
\cite{Boulware:1973my,Boulware:1974sr}. 
This is because due to the non-linearity, $h_{00}$ cannot be the Lagrange 
multiplier field and there appears one extra scalar field. 
After that, there had not been much progress for a long time but recently, remarkable formulations to 
construct ghost free models of massive gravity \cite{deRham:2010ik,deRham:2010kj} and bigravity 
\cite{Hassan:2011hr,Hassan:2011vm,Hassan:2011zd} have been found. 
The bimetric gravity or bigravity has the dynamical metric $f_{\mu\nu}$ and therefore the model is background 
independent. This model includes both of the massless graviton and massive graviton. 

The $F(R)$ gravity extension \cite{Nojiri:2012re,Nojiri:2012zu,Kluson:2013yaa}
has been also well-studied and it has been shown that arbitrary evolution of the expansion of the universe 
can be reproduced. 

\section{New theory of massive spin-two field}

Recently new ghost free interactions called ``pseudo'' linear terms has been proposed 
\cite{Hinterbichler:2013eza} (see also \cite{Folkerts:2011ev}), 
\begin{align}
\label{T11}
\mathcal{L}_{d, n} \sim &\eta^{\mu_{1} \nu_{1} \cdots \mu_{n} \nu_{n}} 
\partial_{\mu_{1}} \partial_{\nu_{1}} h_{\mu_{2} \nu_{2}} \nn
& \cdots \partial_{\mu_{d-1}} \partial _{\nu_{d-1}} h_{\mu_{d} \nu_{d}} h_{\mu_{d+1} \nu_{d+1}} \, .
\end{align} 
Here $\eta^{\mu_{1} \nu_{1} \mu_{2} \nu_{2}} \equiv \eta^{\mu_{1} \nu_{1}} \eta^{\mu_{2} \nu_{2}} 
 - \eta^{\mu_{1} \nu_{2}} \eta^{\mu_{2} \nu_{1}}$, 
$\eta^{\mu_{1} \nu_{1} \mu_{2} \nu_{2} \mu_{3} \nu_{3}} \equiv 
\eta^{\mu_{1} \nu_{1}}\eta^{\mu_{2} \nu_{2}} \eta^{\mu_{3} \nu_{3}} 
 - \eta^{\mu_{1} \nu_{1}}\eta^{\mu_{2} \nu_{3}} \eta^{\mu_{3} \nu_{2}} 
+ \eta^{\mu_{1} \nu_{2}}\eta^{\mu_{2} \nu_{3}} \eta^{\mu_{3} \nu_{1}} 
 - \eta^{\mu_{1} \nu_{2}}\eta^{\mu_{2} \nu_{1}} \eta^{\mu_{3} \nu_{3}} 
+ \eta^{\mu_{1} \nu_{3}}\eta^{\mu_{2} \nu_{1}} \eta^{\mu_{3} \nu_{2}} 
 - \eta^{\mu_{1} \nu_{3}}\eta^{\mu_{2} \nu_{2}} \eta^{\mu_{3} \nu_{1}}$. 
The terms in (\ref{T11}) is linear with respect to $h_{00}$ in the Hamiltonian\footnote{
For example,  
\begin{align}
\eta^{\mu_{1} \nu_{1} \mu_{2} \nu_{2}} h_{\mu_{1} \nu_{1}} h_{\mu_{2} \nu_{2}} 
\sim & h_{00} \left( h_{11} + h_{22} + h_{33} \right) 
+ \mbox{terms not including $h_{00}$} \, , \nn
\eta^{\mu_{1} \nu_{1} \mu_{2} \nu_{2} \mu_{3} \nu_{3}} 
\left( \partial_{\mu_{1}} \partial_{\nu_{1}} h_{\mu_{2} \nu_{2}}\right) h_{\mu_{3} \nu_{3}}
\sim & \left( \partial_1^2 h_{00} \right) \left(h_{22} + h_{33} - 2 h_{23} h_{32} \right) 
+ \cdots \, .\nonumber
\end{align}
}
and there do not appear the terms which include both of $h_{00}$ and $h_{0i}$. 
Therefore the variation of $h_{00}$ gives a constraint for $h_{ij}$ and their conjugate momenta $\pi_{ij}$. 
Then by including the secondary constraint, we can eliminate the ghost and we may obtain a 
power-counting renormalizable model of the massive spin two field, whose Lagrangian density is 
given by
\begin{align}
\label{T13}
\mathcal{L}_{h0} 
= & \frac{1}{2} \eta^{\mu_{1} \nu_{1} \mu_{2} \nu_{2} \mu_{3} \nu_{3}} 
\left( \partial_{\mu_{1}} \partial_{\nu_{1}} h_{\mu_{2} \nu_{2}}\right) h_{\mu_{3} \nu_{3}} \nn
& - \frac{m^2}{2} \eta^{\mu_{1} \nu_{1} \mu_{2} \nu_{2}} h_{\mu_{1} \nu_{1}} h_{\mu_{2} \nu_{2}} \nn
& - \frac{\mu}{3!} \eta^{\mu_{1} \nu_{1} \mu_{2} \nu_{2} \mu_{3} \nu_{3}} 
h_{\mu_{1}\nu_{1}} h_{\mu_{2} \nu_{2}} h_{\mu_{3} \nu_{3}} \nn
& - \frac{\lambda}{4!} \eta^{\mu_{1} \nu_{1} \mu_{2} \nu_{2} \mu_{3} \nu_{3} \mu_{4} \nu_{4}} 
h_{\mu_{1} \nu_{1}} h_{\mu_{2} \nu_{2}} h_{\mu_{3} \nu_{3}} h_{\mu_{4} \nu_{4}} \, .
\end{align}
Here $m$ and $\mu$ are parameters with the dimension of mass and $\lambda$ is a dimensionless parameter.
Therefore the model (\ref{T13}) is power-counting renormalizable and of course free from ghost. 
The model is not, however, really renormalizable. 
This is because the propagator is given by
\begin{align}
\label{T14}
D^m_{\alpha\beta,\rho\sigma} =&  
\frac{1}{2\left( p^2 + m^2 \right)}
\left\{  P^m_{\alpha\rho} P^m_{\beta\sigma} 
+ P^m_{\alpha\sigma} P^m_{\beta\rho} \right. \nn
& \left. - \frac{2}{D-1} P^m_{\alpha\beta} P^m_{\rho\sigma} \right\} \, . 
\end{align}
Here $P^m_{\mu\nu} \equiv \eta_{\mu\nu} + \frac{p_\mu p_\nu}{m^2}$ is a projection operator on mass-shell. 
Due to the projection operator, when $p^2 \to \infty$, the propagator behaves as 
$D^m_{\alpha\beta,\rho\sigma} \sim \mathcal{O}\left( p^2 \right)$ and therefore the model (\ref{T13}) 
is not renormalizable. 

\begin{figure*}[hbp]
\begin{center}

\unitlength=0.45mm

\begin{picture}(200,100)

\thicklines
\qbezier(50,80)(40,80)(30,70)
\qbezier(30,70)(20,60)(10,40)
\qbezier(50,80)(60,80)(70,70)
\qbezier(70,70)(80,60)(90,40)

\put(50,80){\makebox(0,0){\circle*{4}}}

\put(50,90){\makebox(0,0){local maximum}}
\put(50,30){\makebox(0,0){no tachyon for $h_{\mu\nu}$ ($m^2>0$)}}
\put(50,20){\makebox(0,0){stable}}
\put(50,10){\makebox(0,0){$C$ does not roll down}}

\qbezier(150,50)(140,50)(130,60)
\qbezier(130,60)(120,70)(110,90)
\qbezier(150,50)(160,50)(170,60)
\qbezier(170,60)(180,70)(190,90)

\put(150,50){\makebox(0,0){\circle*{4}}}

\put(150,40){\makebox(0,0){local minimum}}
\put(150,30){\makebox(0,0){tachyon for $h_{\mu\nu}$ ($m^2<0$)}}
\put(150,20){\makebox(0,0){unstable}}

\end{picture}

\caption{
Upward convex potential (left) is stable but downward convex potential is unstable, which might be 
counter-intuitive.   
}
\label{fig1}
\end{center}

\end{figure*}
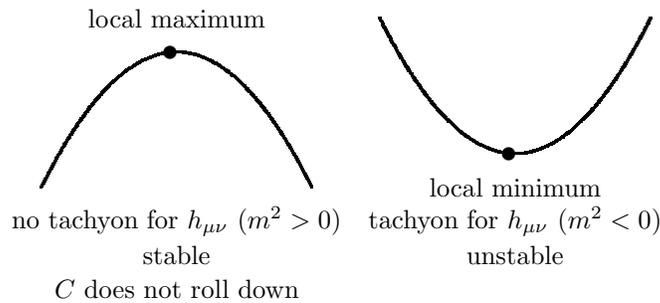

We may consider the classical solution by assuming $h_{\mu\nu} = C \eta_{\mu\nu}$ with a constant $C$. 
Then the action reduces to the following form: 
$S = - \int d^4 x V(C)$. Here $V(C) \equiv - 6m^2 C^2 + 4\mu C^3 + \lambda C^4$. 
Then $C$ can be determined by the condition $V'(C)=0$. 
We should note that when $\mu=\lambda=0$, which corresponds to the Fierz-Pauli model, 
$V(C)$ is unbounded below but this does not generate any inconsistency because $C$ does not propagate 
and therefore does not roll down the potential. 
On the other hand, on the local minimum of the potential ($m^2 < 0$), 
$h_{\mu\nu}$ becomes tachyon and therefore the local minimum corresponds to the instability 
(see Fig.\ref{fig1}). 

In order to show that $C$ is always a constant, we consider the equation of motion given by
\begin{align}
\label{T16}
0 =& \eta^{\mu \nu \mu_{1} \nu_{1} \mu_{2} \nu_{2}} 
\partial_{\mu_{1}} \partial_{\nu_{1}} h_{\mu_{2} \nu_{2}} 
 - m^2 \eta^{\mu \nu \mu_{1} \nu_{1}} h_{\mu_{1} \nu_{1}} \nn
& - \frac{\mu}{2} \eta^{\mu\nu \mu_{1} \nu_{1} \mu_{2} \nu_{2}} 
h_{\mu_{1}\nu_{1}} h_{\mu_{2} \nu_{2}} \nn
& - \frac{\lambda}{3!} \eta^{\mu \nu \mu_{1} \nu_{1} \mu_{2} \nu_{2} \mu_{3} \nu_{3}} 
h_{\mu_{1} \nu_{1}} h_{\mu_{2} \nu_{2}} h_{\mu_{3} \nu_{3} } \, .
\end{align}
When we assume  $h_{\mu\nu} = C \eta_{\mu\nu}$ but $C$ is not a constant, Eq.~(\ref{T16}) 
reduces into the following form:
\be
\label{T17}
0 = \eta^{\mu\nu} \left( 2 \Box C - 3 m^2 C - 3 \mu C^2 + 3 \lambda C^3 \right)
 - 2 \partial^\mu \partial^\nu C\, ,
\ee
which tells that $C$ should be a constant and therefore even if $C$ is on the local maximum 
of the potential, $C$ does not roll down. 
This could be consistent because only massive spin two mode can propagate but there is no propagating scalar 
mode in the model. 

For simplicity, we parametrize $m^2$ and $\mu$ by using new parameters $C_1$ and $C_2$, 
$m^2 = - \frac{\lambda}{3} C_1 C_2$ and $\mu = - \frac{\lambda}{3} \left( C_1 + C_2 \right)$. 
Then the solutions of the equation $V'(C)=0$ are given by $C=0,\, C_1,\, C_2$ and we find 
$V\left(C_1\right) = \frac{\lambda}{3} C_1^3\left( - C_1 + 2 C_2 \right)$ and 
$V\left(C_2\right) = \frac{\lambda}{3} C_2^3\left( - C_2 + 2 C_1 \right)$. 
If we naively couple the model with gravity, $V(C)$ plays the role of the cosmological constant. 
Then we might obtain the accelerating expansion of the universe, which could be the de Sitter 
space-time, if $V>0$.
On the other hand, $V<0$ could correspond to the anti-de Sitter space-time. 
When $\lambda>0$, there appear two unstable anti-de Sitter space-time. 
On the other hand, when $\lambda<0$,  two cases appear. 
In one case, there are two solutions corresponding to the unstable anti-de Sitter 
space-time and the stable de Sitter space-time. 
In another case, one solution is 
the stable anti-de Sitter space-time and another is the unstable anti-de Sitter space-time. 
When we consider the supersymmetric model, 
if $E>0$, there might occurs the breaking of supersymmetry.
The relation between $C_{1,2}$ and the corresponding space-time and the 
stability of the solutions is summarized in Table.\ref{table1}. 
We should also note that $C_1$ and $C_2$ are given by 
$C_{1}=\frac{-3 \mu + \sqrt{9 \mu^2+12m^2 \lambda}}{2 \lambda}$ and 
$C_{2}=\frac{-3 \mu - \sqrt{9 \mu^2+12m^2 \lambda}}{2 \lambda}$.

\begin{table*}[hbp]
\label{table1}

\caption{The relation between $C_{1,2}$ and the corresponding space-time and the 
stability of the solutions}
\begin{center}

\begin{tabular}{|c|c|c|c|} \hline 
& $0<\lambda $ & 
$- \frac{2\mu^2}{3m^2} < \lambda < 0$ 
& $ - \frac{3\mu^2}{4m^2} < \lambda <- \frac{2\mu^2}{3m^2}$ \\ \hline
de Sitter & no solution & $C_2$ (stable)& no solution \\ \hline
Anti-de Sitter & \parbox{4.5cm}{$C_1$ (unstable) \\ $C_2$ (unstable)} 
& $C_1$ (unstable) & \parbox{4.5cm}{$C_1$ 
(unstable) \\ $C_2$ (stable)} \\ \hline
\end{tabular}

\end{center}

\end{table*}

\section{New bigravity}

If we couple the model of massive spin two field with the gravity, the model might be regarded 
as a new bigravity, 
\begin{align}
\label{T19}
S =& \int d^4 x \sqrt{-g} \left\{ 
\frac{1}{2} g^{\mu_{1} \nu_{1} \mu_{2} \nu_{2} \mu_{3} \nu_{3}} 
\nabla_{\mu_{1}} \nabla_{\nu_{1}} h_{\mu_{2} \nu_{2}} h_{\mu_{3} \nu_{3}}
\right. \nn
&  - \frac{1}{2} m^2 g^{\mu_{1} \nu_{1} \mu_{2} \nu_{2}} h_{\mu_{1} \nu_{1}} h_{\mu_{2} \nu_{2}} \nn
& \left. - \frac{\mu}{3!} g^{\mu_{1} \nu_{1} \mu_{2} \nu_{2} \mu_{3} \nu_{3}} 
h_{\mu_{1}\nu_{1}} h_{\mu_{2} \nu_{2}} h_{\mu_{3} \nu_{3}} \right. \nn
& \left. 
 - \frac{\lambda}{4!} g^{\mu_{1} \nu_{1} \mu_{2} \nu_{2} \mu_{3} \nu_{3} \mu_{4} \nu_{4}} 
h_{\mu_{1} \nu_{1}} h_{\mu_{2} \nu_{2}} h_{\mu_{3} \nu_{3}} h_{\mu_{4} \nu_{4}} \right\} 
\, .
\end{align}
Here $h_{\mu\nu}$ is not the perturbation in $g_{\mu\nu}$ but 
$h_{\mu\nu}$ is a field independent of $g_{\mu\nu}$.

Naively if we work in the local Lorentz frame, we might expect that there does not appear any ghost. 
Buchbinder et al. \cite{Buchbinder:1999be,Buchbinder:1999ar}, however, have shown that even in case of 
the Fierz-Pauli model, consistent theory should be
\begin{align}
\label{T20}
S=& \int d^D x\sqrt{-g} \left\{ \frac{1}{4} \nabla_\mu h \nabla^\mu h
 -\frac{1}{4} \nabla_\mu h_{\nu\rho} \nabla^\mu h^{\nu\rho} \right. \nn
& -\frac{1}{2} \nabla^\mu h_{\mu\nu} \nabla^\nu h 
+\frac{1}{2} \nabla_\mu h_{\nu\rho} \nabla^\rho h^{\nu\mu} 
+\frac{\xi}{2D} R h_{\mu\nu} h^{\mu\nu} \nn
& \left. +\frac{1-2\xi}{4D} R h^2
 - \frac{m^2}{4} h_{\mu\nu} h^{\mu\nu} + \frac{m^2}{4} h^2
\right\} \, .
\end{align}
Furthermore the background space-time should be the Einstein space where 
$R_{\mu\nu}=\frac{1}{D}g_{\mu\nu}R$.\footnote{
The inconsistence when the Fierz-Pauli model couples with gravity was found in \cite{Aragone:1979bm}. 
} 
In case of interacting model in this report, we have shown that the consistent model is given by 
\begin{align}
\label{T21}
S=& \int d^4x \sqrt{-g}\left\{ \frac12 \nabla_\mu h \nabla^\mu h 
 -\frac12 \nabla_\mu h_{\nu \rho} \nabla^\mu h^{\nu \rho} \right. \nn
& - \nabla^\mu h_{\mu \nu} \nabla^\nu h + \nabla_\mu h_{\nu \rho } \nabla^\rho h^{\nu \mu}
+\frac{\xi}{4} R h_{\alpha \beta} h^{\alpha \beta} \nn
& +\frac{1-2\xi}{8} R h^2
+\frac{m^2}{2} g^{\mu_1 \nu_1 \mu_2 \nu_2 } h_{\mu_1 \nu_1} 
h_{\mu_2 \nu_2} \nn
& -\frac{\mu}{ 3!} g^{\mu_1 \nu_1 \mu_2 \nu_2 \mu_3 \nu_3 } h_{\mu_1 \nu_1} 
h_{\mu_2 \nu_2} h_{\mu_3 \nu_3} \nn
& \left. 
 - \frac{\lambda}{4!} \eta^{\mu_{1} \nu_{1} \mu_{2} \nu_{2} \mu_{3} \nu_{3} \mu_{4} \nu_{4}} 
h_{\mu_{1} \nu_{1}} h_{\mu_{2} \nu_{2}} h_{\mu_{3} \nu_{3}} h_{\mu_{4} \nu_{4}}
\right\}\, .
\end{align}
In Eq.~(\ref{T21}), the changes from (\ref{T19}) are only for quadratic terms as in 
the Fierz-Pauli model. 
Therefore we may consider (anti-)de Sitter (-Schwarzschild or Kerr) space-time as 
exact solutions. 

By assuming $h_{\mu\nu} = C g_{\mu\nu}$ with a constant $C$, the action (\ref{T21}) can be rewritten as 
\begin{align}
\label{T22}
S =& - \int d^4x \sqrt{-g} V(C) +\frac{1}{2\kappa^2} \int d^4x\sqrt{-g} R\, , \nn
V(C) = & - \left\{ 6m^2 + \left( 2-3\xi \right)R \right\} C^2 
+ 4\mu C^3 + \lambda C^4 \, , \nn
S =& \left\{ \left( 2-3\xi \right) C^2+\frac{1}{2\kappa^2} \right\} 
\int d^4 x \sqrt{-g} \left[ R -2 \Lambda_\mathrm{eff} \right] \, .
\end{align}
Then the effective mass $M$ is given by $M^2 \equiv m^2 -2\mu C -\lambda C^2$ and 
the effective cosmological constant by 
$\Lambda_\mathrm{eff} \equiv \frac{\kappa^2 \left(-6m^2 C^2 + 4\mu C^3 
+ \lambda C^4 \right)}{2\kappa^2 C^2 \left(2-3 \xi \right) +1 }$. 
Then we obtain $R = 4 \Lambda_\mathrm{eff}$, which tells 
${V_0}'(C) = 4C \left\{ -2\mu \zeta C^3 + \left(\lambda + 6\zeta m^2\right) C^2 
+3\mu C -3m^2 \right\} =0$. 
Here $\zeta \equiv \kappa^2 \left( 2-3\xi \right)$. 
Besides a trivial solution $C=0$, the non-trivial solutions can be obtained by defining 
$C=4 x+\frac{\lambda+ 6\zeta m^2 }{6\mu \zeta }$, 
$p= -\frac13 \left\{ \left( \frac{\lambda + 6\zeta m^2}{2\mu \zeta} \right)^2 
+\frac{9}{2\zeta} \right\}$, 
$q= \frac{2}{27} \left( \frac{\lambda + 6\zeta m^2}{2\mu \zeta} \right)^3 
 -\frac{\lambda}{4\mu \zeta^2}$, and $\omega \equiv \e^{i 2\pi /3}$. 
Then the explicit expressions of solutions are given by
\begin{align}
\label{T26}
x= &\omega^k \sqrt[3]{-\frac{q}{2} + \sqrt{\left( \frac{q}{2} \right)^2 
+ \left( \frac{p}{3} \right)^3} } \nn
&+ \omega^{3-k} \sqrt[3]{-\frac{q}{2} - \sqrt{\left( \frac{q}{2} \right)^2 
+ \left( \frac{p}{3} \right)^3} } \, , \quad k=1,2,3\, .
\end{align} 
and the determinant is given by 
$D= -27q^2 -4 p^3 =-2^2 \cdot 3^3 \left\{ \left( \frac{q}{2} \right)^2 
+ \left( \frac{p}{3} \right)^3 \right\}$
Then except the case $q=p=0$, we find, 
\begin{enumerate}
\item\label{case1} $D>0$ There are three different real solutions. 
\item\label{case2} $D<0$ There is only one real solution. 
\item\label{case3} $D=0$ There are three real solutions but two of them are degenerate 
with each other. 
\end{enumerate}
We should note that the stability of the solution is related with the Higuchi bound \cite{Higuchi:1986py}. 

\section{Summary}

We propose a new theory describing massive spin two field and consider the coupling of the theory with 
gravity, which may be a new kind of bimetric gravity or bigravity. 
Then we find the massive spin two field plays the role of the cosmological constant. 
We have also shown that
\begin{itemize}
\item The conditions of no ghost is not changed from those in the Fierz-Pauli case. 
\item There could be the accelerating expansion of the universe (inflation or dark energy). 
\item The (anti-)de Sitter-Schwarzschild (Kerr) space-time is an exact solution.
\end{itemize}
It could be interesting to consider if the particle corresponding to the massive spin two field 
can be dark matter.

Because the (ant-) de Sitter-Schwarzschild (Kerr) black hole space-time is an exact solution, it 
could be interesting to investigate the entropy of the black hole. 
As a tentative result, we have found that the entropy 
would not be changed from the case of the Einstein gravity, whose situation is different from the 
Hassan-Rosen bigravity.case \cite{Katsuragawa:2013bma,Katsuragawa:2013lfa}, where 
the entropy is the sum of the contributions from two metric sectors 
corresponding to $g_{\mu\nu}$ and $f_{\mu\nu}$.

\section*{Acknowledgments} 

The author is indebted to Ohara and Akagi for the collaborations and Odintsov and Katsuragawa 
for the discussions. 
He thanks for the hospitality when he stayed in Tomsk State Pedagogical University, 
especially Rector Obukhov, Buchbinder, Epp, Makarenko, and Osetrin.  
The work is supported by the JSPS Grant-in-Aid for Scientific 
Research (S) \# 22224003 and (C) \# 23540296 (S.N.). 

\end{multicols}

\end{document}